\documentclass{PoS}

\usepackage{amsmath}
\usepackage{epsfig}
\usepackage{graphicx}
\newlength{\bredde}
\def\slash#1{\settowidth{\bredde}{$#1$}\ifmmode\,\raisebox{.15ex}{/}
\hspace*{-\bredde} #1\else$\,\raisebox{.15ex}{/}\hspace*{-\bredde} #1$\fi}

\newcommand{\la}{\langle}
\newcommand{\ra}{\rangle}

\def\amp{{\mathcal{M}}}

\def\beq{\begin{equation}}
\def\bea{\begin{eqnarray}}
\def\eeq{\end{equation}}
\def\eea{\end{eqnarray}}

\def\caliphi{\varphi}
\def\caliD{{\mathcal{D}}}
\def\la{\langle}
\def\ra{\rangle}

\title{Study of the Higgs-Yukawa theory in the strong-Yukawa coupling
  regime \thanks{{Preprint number: DESY 11-219}}}

\ShortTitle{Strong-Yukawa coupling}

\author{John~Bulava$^{a}$, Philipp~Gerhold$^{b,c}$, 
     George~W.-S.~Hou$^{d}$, Karl~Jansen$^{c}$, Bastian~Knippschild$^{d,e}$, 
     \speaker{C.-J.~David~Lin}$^{f,g,\dagger}$, Kei-Ichi~Nagai$^{h}$, 
     Attila~Nagy$^{b,c}$, Kenji~Ogawa$^{f}$, Brian~Smigielski$^{d}$\\
$^{a}$ CERN, Physics Department, 1211 Geneva 23, Switzerland\\
$^{b}$ Institut f\"{u}r Physik, Humboldt-Universit\"{a}t zu Berlin, D-12489 Berlin, Germany\\
$^{c}$ NIC, DESY, D-15738 Zeuthen, Germany\\
$^{d}$ Department of Physics, National Taiwan University, Taipei 10617, Taiwan\\
$^{e}$ Institut f\"{u}r Kernphysik, University of Mainz, D-55099 Mainz, Germany\\
$^{f}$ Institute of Physics, National Chiao-Tung University, Hsinchu  300, Taiwan\\
$^{g}$ Division of Physics, National Centre for Theoretical Sciences, Hsinchu 300, Taiwan\\
$^{h}$ Kobayashi-Maskawa Institute, Nagoya University, Nagoya 464-8602, Japan\\ \ \\
       $^{\dagger}$E-mail: \email{dlin@mail.nctu.edu.tw}}

\abstract{In this article, we present an ongoing lattice study of the Higgs-Yukawa model, in the 
          regime of strong-Yukawa coupling, using overlap
          fermions.  We investigated the phase structure
          in this regime by computing the Higgs vacuum expectation
          value, and by exploring the finite-size scaling
          behaviour of the susceptibility corresponding to the magnetisation.  Our preliminary results indicate the
          existence of a second-order phase transition
          when the Yukawa coupling becomes large enough, at which the
          Higgs vacuum expectation value vanishes and the susceptibility diverges.}

\FullConference{ The XXIX International Symposium on Lattice Field Theory - Lattice 2011\\
July 10-16, 2011\\
Squaw Valley, Lake Tahoe, California}

\begin{document}

\section{Introduction}
In recent years, there have been interests in the possible existence
of heavy extra-generation fermions (mass $\ge$ 600 GeV) beyond the 
standard model (SM).  Such heavy fermions are a consequence of
strong-Yukawa couplings.  Their presence in nature remains to be
examined by experimental data that will be collected at the LHC.
An important consequence of a 4th fermion generation is the
substantial enhancement the amount of CP violation~\cite{Hou:2008xd}. 
Large bare values of the Yukawa coupling may also lead to the
formation of bound states which can replace the role of the Higgs
boson in unitarising the WW scattering process~\cite{Holdom:2006mr,Holdom:2011fc,Hung:2010xh}.  Such a
scenario is clearly of nonperturbative nature and motivates the use
of lattice field theory as a first-principle and nonperturbative
tool for this research  avenue.  Lattice investigations 
at small and moderate values of the bare Yukawa coupling showed~\cite{Gerhold:2010wv}
that the lower Higgs boson mass bound is strongly affected by the
presence of a heavy 4th fermion generation when compared to results
using a physical value of the top quark mass~\cite{Gerhold:2009ub,Gerhold:2010bh}.
Still, in these simulations no signs of bound states were observed, as expected.

\smallskip

However, lattice simulations have also revealed the existence of an interesting
phase structure of the model at large values of the Yukawa 
coupling~\cite{Hasenfratz:1989jr,Hasenfratz:1988vc,Hasenfratz:1991it,Polonyi:1988jq,Lee:1989xq,Lee:1989mi,Bock:1991bu,Bock:1991da}.
These simulations, performed around 1990, were lacking
an exact chiral symmetry on the lattice and the results of these works
are therefore not easy to interpret and to connect to the SM.
Recently, exact lattice chiral symmetry~\cite{Luscher:1998pqa} were
established and, in fact, lattice simulations employing this lattice
chiral symmetry confirmed the phase structure at strong bare Yukawa
coupling~\cite{Gerhold:2007yb,Gerhold:2007gx} as found in the earlier studies.

\smallskip

In this work we further explore the phase structure of a lattice chirally invariant Higgs-Yukawa model at large values of the
Yukawa coupling. Our main aim is to start a systematic investigation,
whether the phase transitions between the symmetric phase with
vanishing vacuum expectation value (VEV) $v=0$ and the broken phase with $v>0$ are governed by critical exponents
that differ from the (Gaussian) one of the SM. This is a
highly non-trivial question. In particular, it can 
be shown that the lattice Higgs-Yukawa model in the limit of infinite bare Yukawa
coupling reduces to a pure scalar non-linear 
$\sigma$-model~\cite{Hasenfratz:1989jr,Hasenfratz:1988vc,Hasenfratz:1991it,Gerhold:2007yb,Gerhold:2007gx} with again Gaussian critical exponents. Hence, the 
most interesting and important question is, whether at large but
finite value of the Yukawa-coupling a new and non-trivial 
universality class emerges. We emphasise that a quantitative
determination of critical exponents 
of the phase transitions in the strong Yukawa coupling region
using chiral invariant lattice fermions was never attempted before.
The above potential 
of nonperturbative physics at large Yukawa couplings motivates clearly such an investigation.

\section{Simulation details}

The discretisation of the scalar
field theory leads to the action (with lattice spacing $a$ set to $1$)

\beq
\label{eq:naive_scalar_action}
 S_{\caliphi} = - \sum_{x,\mu} \caliphi_{x}^{\alpha} \caliphi^{\alpha}_{x + \hat{\mu}} + 
       \sum_{x} \left [ \frac{1}{2} (2d+ m_{0}^{2}) \caliphi_{x}^{\alpha}\caliphi_{x}^{\alpha}
       + \frac{1}{4}\lambda_{0} (\caliphi_{x}^{\alpha}\caliphi_{x}^{\alpha})^{2} \right ] ,
\eeq
where $\alpha$ labels the four components of the scalar fields, 
$d$ is the number of the space-time dimensions, $m_{0}$ is the bare mass and $\lambda_{0}$
is the bare quartic self-coupling.  For practical lattice simulations, it is convenient
to perform the change of variables
\beq
\label{eq:scalar_change_of_variables}
 \caliphi = \sqrt{2\kappa} \phi ,\mbox{ }\mbox{ }
 m^{2}_{0} = \frac{1 - 2 \hat{\lambda} - 2 d \kappa}{\kappa},\mbox{ }\mbox{ }
 \lambda_{0} = \frac{\hat{\lambda}}{\kappa^{2}} ,
\eeq
where $\kappa$ is the hopping parameter. This renders the lattice scalar field theory
to the Ising form
\beq
\label{eq:scalar_action_ising_form}
  S_{\phi} = - 2\kappa 
      \sum_{x,\mu} \phi^{\alpha}_{x} \phi^{\alpha}_{x + \hat{\mu}}
       + \sum_{x} \left [ \phi^{\alpha}_{x}\phi^{\alpha}_{x} + 
        \hat{\lambda}(\phi^{\alpha}_{x}\phi^{\alpha}_{x}-1)^{2}\right ] ,
\eeq
which is more suitable for exploring the phase structure.

\smallskip

In this work, we use the overlap operator $\caliD^{({\mathrm{ov}})}$ in the lattice fermion action
\beq
\label{eq:fermion_action}
 S_{F} = \bar{\Psi} \amp \Psi , \mbox{ }{\mathrm{where}}\mbox{ }\mbox{ }
  \amp = \caliD^{({\mathrm{ov}})} 
    + P_{+} \mbox{ }\Phi^{\dagger}\mbox{ }\mbox{ }{\mathrm{diag}}(y_{t^{\prime}},y_{b^{\prime}})\mbox{ }\hat{P}_{+}
    + P_{-} \mbox{ }\mbox{ }{\mathrm{diag}}(y_{t^{\prime}},y_{b^{\prime}})\mbox{ }\Phi\mbox{ }\hat{P}_{-} ,
\eeq
with
\beq
\label{eq:Phi_and_Psi}
 \Psi = \left ( \begin{array}{c} t^{\prime} \\ b^{\prime} \end{array} \right ), \mbox{ }
   {\mathrm{and}}\mbox{ }\mbox{ } 
 \Phi = \left ( \begin{array}{c} \phi^{2} + i \phi^{1} \\ \phi^{0} - i \phi^{3}\end{array} \right ) .
\eeq
The chiral projectors are defined as
\beq
 P_{\pm} = \frac{1 \pm \gamma_{5}}{2},\mbox{ }\mbox{ }
 \hat{P}_{\pm} = \frac{1 \pm \hat{\gamma}_{5}}{2}, \mbox{ }\mbox{ }
 \hat{\gamma}_{5} = \gamma_{5} \left ( 1 -\frac{1}{\rho} \caliD^{({\mathrm{ov}})} \right ),
\eeq
where $\rho$ is the radius of the circle of eigenvalues in the complex plane of the free overlap
operator.  We also set $y_{t^{\prime}} = y_{b^{\prime}} = y$, 
to ensure that the fermion determinants are positive definite.

\smallskip

Our simulations have been performed at two $\kappa$ values, $\kappa = 0.00$ and $0.06$, with the bare
Yukawa coupling $y$ in the range between $14$ and $25$.    The bare scalar quartic coupling
$\hat{\lambda}$ is fixed to infinity, which results in the largest
possible Higgs mass~\cite{Gerhold:2010wv,Gerhold:2009ub,Gerhold:2010bh}.  We already accumulated data with reasonable
statistics for the volumes $8^{3}\times 16$, $12^{3}\times 24$ and $16^{3}\times 32$.  In generating
dynamical scalar field configurations, we use the polynomial HMC (pHMC) algorithm~\cite{Frezzotti:1997ym}, treating the 
weight factor as an observable~\cite{Gerhold:2010wy}.  We have found
that high degrees of 
polynomials ($\sim 180$ for $16^{3}\times 32$ lattices) 
are necessary in order to obtain reasonable statistical accuracy for
the weight factor.  For each set of the bare couplings, we carry out 1000 pHMC trajectories to
precondition the fermion matrix and to thermalise the simulation.  Our
measurements are then performed on $\sim 2000$ thermalised trajectories.

\section{The scalar vacuum expectation value}
We first measure the scalar VEV to probe the phase structure.  Our simulations have been performed 
without external sources, therefore a naive computation of the VEV
would always lead to vanishing results
in finite volume.  In this work, we follow the procedure in 
Refs.~\cite{Hasenfratz:1989ux,Hasenfratz:1990fu,Gockeler:1991ty} to project the scalar fields on
the direction of magnetisation
\beq
\label{eq:mag}
m = \frac{1}{V_{4}} \left ( \sum_{\alpha,x}
  |\phi_{x}^{\alpha}|^{2} \right )^{1/2} \mbox{ }\mbox{ } (V_{4}
\mbox{ }{\mathrm{is}}\mbox{ }{\mathrm{the}}\mbox{ }4{-}{\mathrm{dimensional}}\mbox{ }{\mathrm{volume}}),
\eeq
then compute the VEV.   This ``projected'' VEV coincides with the scalar VEV in the infinite-volume 
limit, and its introduction in finite volume is equivalent to coupling the 
scalar fields to external sources~\cite{Gockeler:1991ty}.

\smallskip

\begin{center}
\begin{figure}[t]
\vspace{-5cm}
\includegraphics[width=15cm, height=15cm,angle=0]{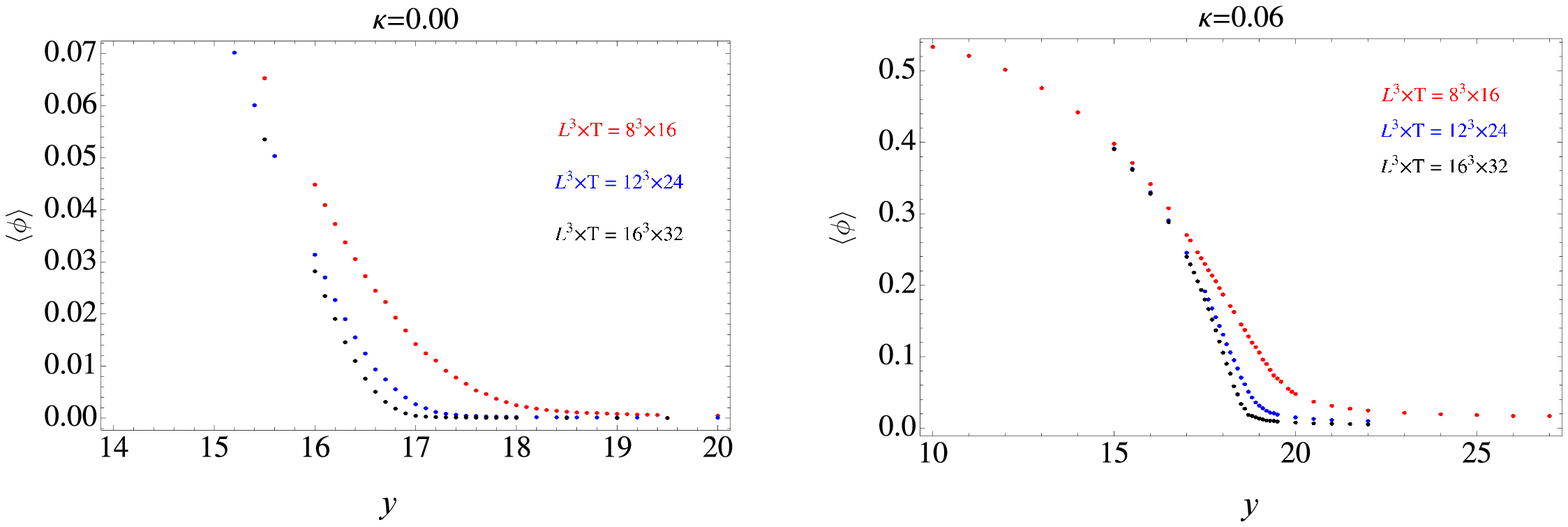}
\vspace{-5cm}
\caption{\label{fig:vev}Bare scalar VEV at two $\kappa$ values with $\hat{\lambda} = \infty$.}
\end{figure}
\end{center}

The results for the bare VEV, computed from the above projection procedure, 
are shown in Fig.~\ref{fig:vev}\footnote{We are currently computing
the Goldstone wavefunction renormalisation to obtain the renormalised Higgs VEV~\cite{Luscher:1988uq}.}.  
It is clear that there is a phase transition when $y$ becomes large,
at which
the system enters a symmetric phase.  These plots also indicate
that the value of $y=y_{{\mathrm{crit}}}$ at which the phase transition occurs grows with $\kappa$.  This
agrees with the qualitative predictions from the strong-coupling expansion~\cite{Abada:1990ds}, the
large-$N_{f}$ expansion~\cite{Gerhold:2007yb} and an exploratory numerical study using overlap 
fermions~\cite{Gerhold:2007gx}.

\smallskip

\section{Finite-size scaling of the magnetisation susceptibility}
In order to determine the order of the phase transition observed in the last section, we 
investigate the finite-size scaling behaviour of the susceptibility
corresponding to the magnetisation. It is defined as
\beq
\label{eq:susceptibility_definition}
\chi = V_{4} \left ( \la m^{2} \ra 
            -  \la m \ra  
               \la m \ra  \right ) ,
\eeq
where $V_{4}$ and $m$ are defined in Eq.~(\ref{eq:mag}).  This
quantity diverges at the critical points in the infinite-volume limit.
Finite-size effects result in the crossover from this bulk behaviour to
the finite-volume scaling behaviour in lattice calculations.
In the vicinity of a would-be second-order phase transition, the solution of the renormalisation group equation
(RGE) in finite volume predicts~\cite{ZinnJustin:2002ru}
\beq
 \chi L_{s}^{-\gamma / \nu} = g(t L_{s}^{1/\nu}), \mbox{ }{\mathrm{with}}\mbox{ }\mbox{ }
  t = (y/(y_{{\mathrm{crit}}}-A_{4}/L^{b}_{s}) - 1),
\label{eq:sus_scaling}
\eeq
where $g$ is a universal function, $L_{s}$ is the spatial volume,
$y_{{\mathrm{crit}}}$ is the critical Yukawa coupling
at which the phase transition occurs in the infinite-volume limit,
$A_{4}$ is a phenomenological parameter,
$\nu$ and $\gamma$ are the universal critical exponents (anomalous dimensions),
and $b$ is the shift exponent~\cite{Fisher:1972zza}.

\smallskip

\begin{center}
\begin{figure}[t]
\vspace{-1cm}
\includegraphics[width=7.5cm, height=7cm,angle=0]{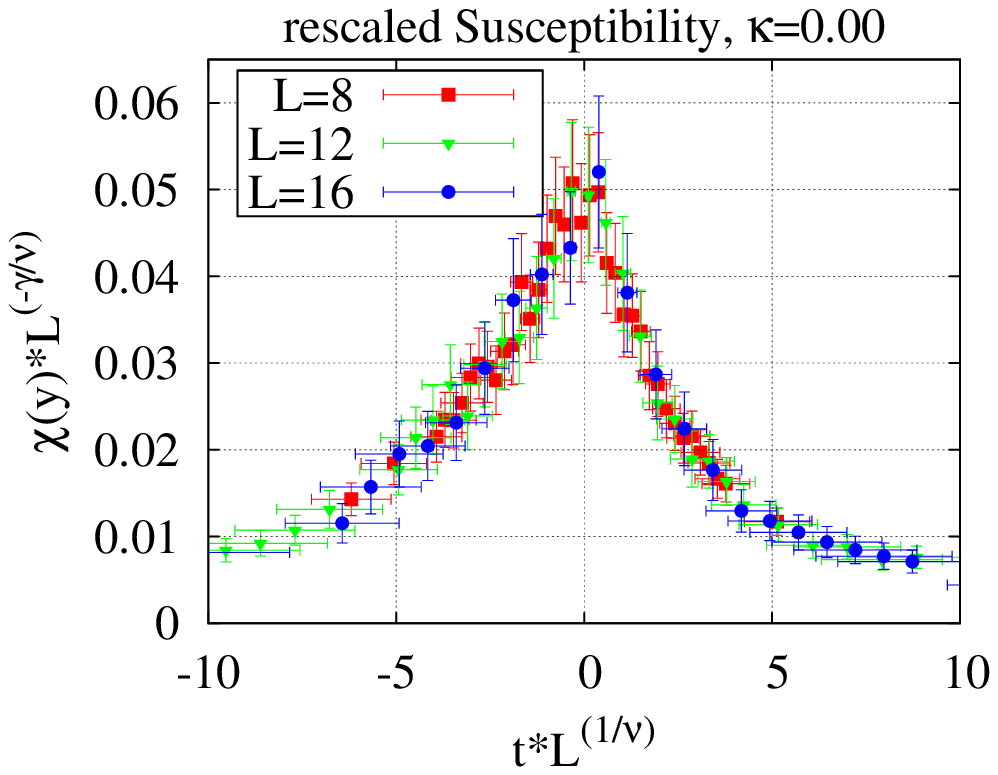}
\hspace{-0.5cm}
\includegraphics[width=7.5cm,
height=7cm,angle=0]{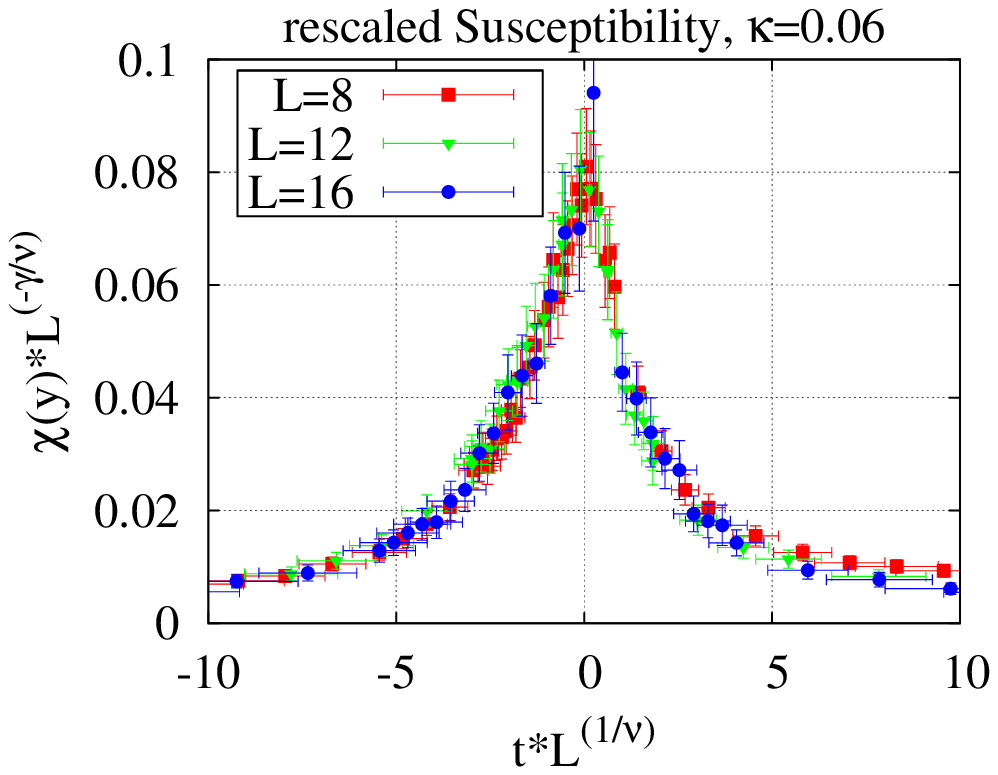}
\vspace{-0.3cm}
\caption{\label{fig:sus_scaling}The scaling behaviour of susceptibility at two $\kappa$ values with $\hat{\lambda} = \infty$.}
\end{figure}
\end{center}

For each of the two $\kappa$ values in our simulations, we perform a simultaneous fit of the
data for the susceptibility at all volumes, to the partly-empirical formula~\cite{Jansen:1989gd}
\beq
\label{eq:chi_FV_fit}
 \chi = A _{1} \left \{ L_{s}^{-2/ \nu} + A_{2,3} 
     \left ( y - y_{{\mathrm{crit}}} -A_{4}/L^{b}_{s} \right )^{2} \right \}^{-\gamma / 2} ,
\eeq
where $A_{1,2,3,4}$ are unknown phenomenological coefficients.  They
are determined, together with $\nu$, $\gamma$, $y_{{\mathrm{crit}}}$
and $b$, from the fits.  For our best procedure, 
the fit ranges of $y$ are $(14.5, 19.5)$ for $\kappa = 0$, and
$(14,22)$ for $\kappa = 0.06$.  The extracted $y_{{\mathrm{crit}}}$,
$\gamma$, $\nu$ and $b$ are presented in
Table~\ref{tab:chi_FV_scaling}.  For comparison with the scaling
behaviour in the regime where the Yukawa couplings are weak and perturbative,
we also list the predictions from the
mean-field calculation in the $O(4)$ scalar model.  
We use these fit results to obtain the susceptibility
as a function of $y$ according to Eq.~(\ref{eq:chi_FV_fit}).  This is
plotted with our data points in Fig.~\ref{fig:sus}.   We also use
the same fit results to construct $\chi L_{s}^{-\gamma / \nu}$ and $t L_{s}^{1/\nu}$, 
to examine the finite-size scaling behaviour of
Eq.~(\ref{eq:sus_scaling}).  The outcome of this test is shown in 
Fig.~\ref{fig:sus_scaling}. 
\begin{center}
\begin{figure}[t]
\vspace{-0.7cm}
\includegraphics[width=7cm, height=6cm,angle=0]{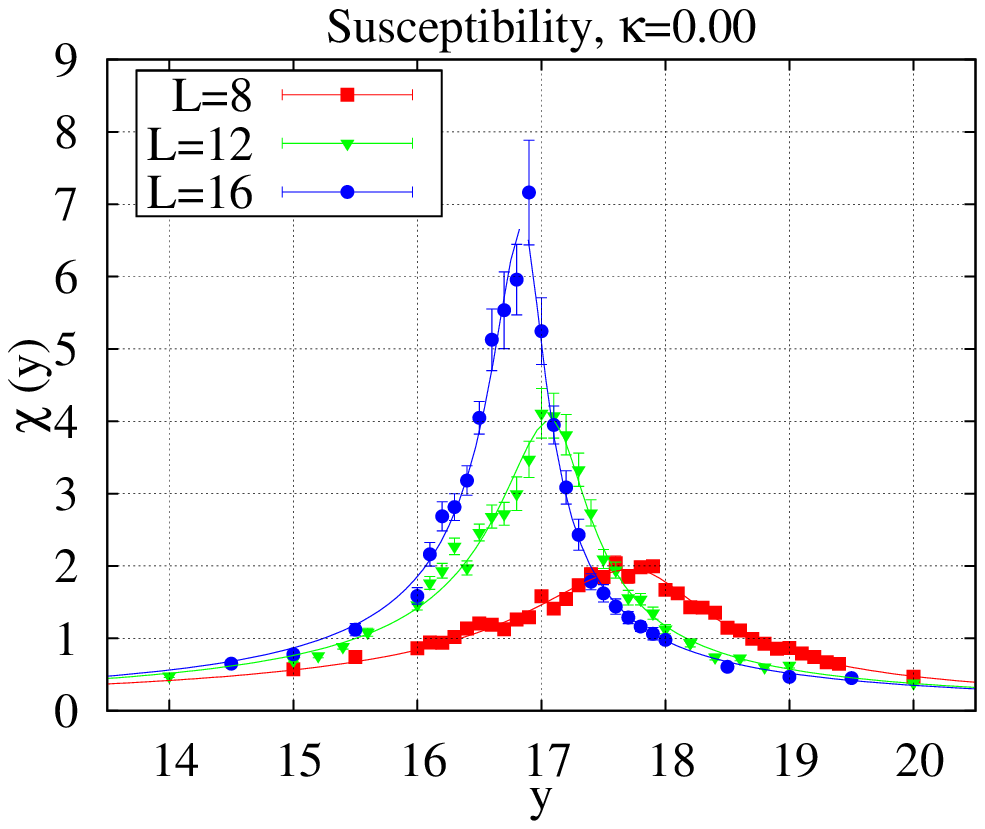}
\hspace{-0.5cm}
\includegraphics[width=7cm,height=6cm,angle=0]{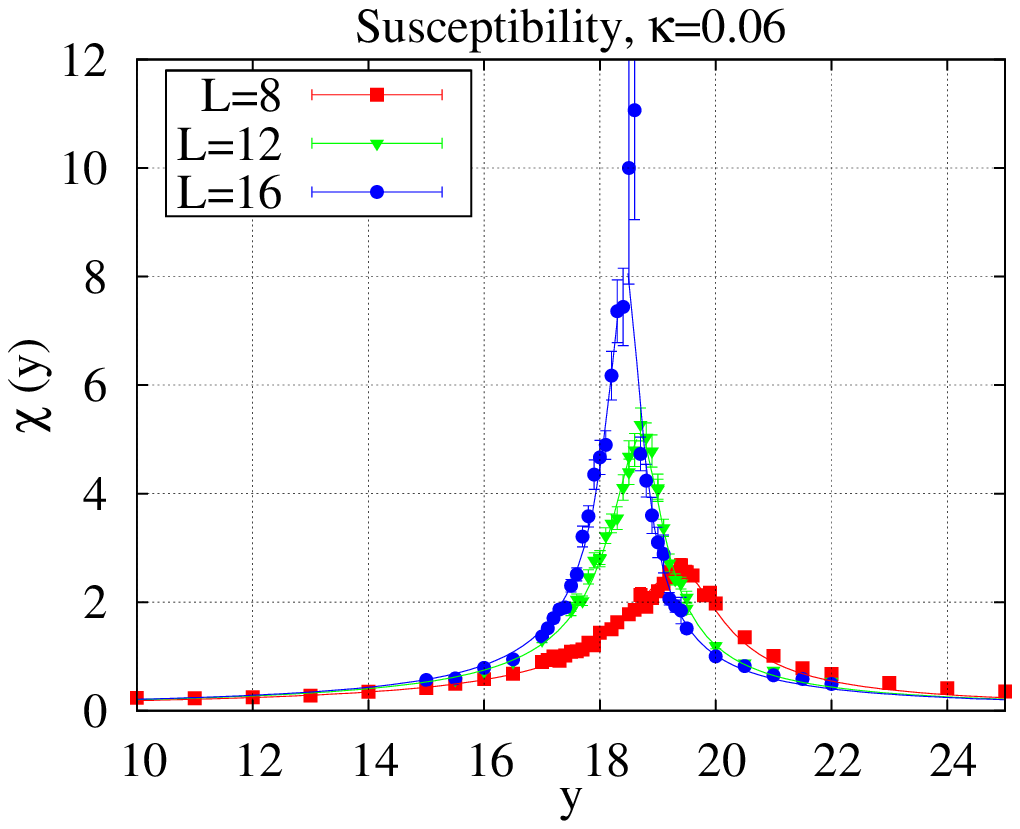}
\vspace{-0.2cm}
\caption{\label{fig:sus}Susceptibility at two $\kappa$ values at various volumes with $\hat{\lambda} = \infty$.}
\end{figure}
\vspace{-0.5cm}
\end{center}
\begin{center}
\vspace{-0.2cm}
\begin{table}[b]
\begin{center}
\begin{tabular}{c|c|c|c}
\hline
    & $\mbox{ }\mbox{ }\kappa = 0.00\mbox{ }\mbox{ }$ & $\mbox{ }\mbox{ }\kappa = 0.06\mbox{ }\mbox{ }$ & O(4) scalar model\\
\hline
$y_{{\mathrm{crit}}}$ & $16.57\pm 0.06$ & $18.11\pm 0.06$ & N/A\\
\hline
$\gamma$ & $1.02\pm 0.02$ & $1.08\pm 0.01$ & 1\\
\hline
$\nu$ & $0.57\pm 0.03$ & $0.66\pm 0.02$ & 0.5\\
\hline
$b$   &  $2.05\pm 0.20$  &  $2.04\pm 0.20$ &  N/A\\
\hline
\end{tabular}
\end{center}
\caption{\label{tab:chi_FV_scaling}The critical Yukawa coupling $y_{{\mathrm{crit}}}$, the critical exponents $\gamma$, $\nu$, 
and the shift exponent $b$ determined from the best fits to the finite-volume scaling function. The errors are statistical only.
Predictions from the mean-field calculation in the $O(4)$ scalar model are also listed.}
\end{table}
\end{center}
\vspace{-1.5cm}
\smallskip

Our data, as presented in Figs.~\ref{fig:sus} and \ref{fig:sus_scaling}, establish evidence for the existence of a second-order
phase transition in the strong-Yukawa coupling regime.  To further investigate the nature of this strong-Yukawa symmetric
phase, we have to perform more detailed studies on the critical
exponents and the spectrum. 
As demonstrated by the results collected in
Table~\ref{tab:chi_FV_scaling}, $\gamma$ is almost consistent with the corresponding
mean-field prediction for the $O(4)$ scalar model, while $\nu$ exhibits a
more significant deviation from it.  By varying the fit ranges in
$y$ in a reasonable interval for the above finite-size scaling analysis, the shifts in $y_{{\mathrm{crit}}}$ and the critical 
exponent $\gamma$ are not statistically distinguishable.  On the other hand, $\nu$ can change by $\sim 11\%$ and become almost 
consistent with the value in the $O(4)$ scalar model.   We have also
tried incorporating logarithmic volume effects in
Eq.~(\ref{eq:chi_FV_fit}) by replacing the coefficient $A_{4}$ with
$A_{4} [1 + c\mbox{ }{\mathrm{Log}}(L_{s})]$ ($c$ is an unknown parameter).
This procedure leads to no statistically significant variation of the
critical exponents, $\nu$ and $\gamma$.
\section{Summary and outlook}
In this article, we present an ongoing numerical investigation of the phase structure of the strong-Yukawa model on the lattice.
Using overlap fermions which respect exact lattice chiral symmetry,
and performing simulations at various volumes, enable us to explore
the details of the phase structure.  From our computation of the scalar VEV and the study of the finite-size scaling
behaviour of the magnetisation susceptibility, we obtain strong
evidence that there exists a symmetric phase in the strong-Yukawa coupling regime,
and that the transition between this phase and the broken phase is of
second-order nature. At the moment, we cannot determine if
the critical exponents for this phase transition are different from those in the weak-Yukawa regime.  

\smallskip

We are currently generating large lattices ($24^{3}\times 48$) which will allow us to have more precise extraction of the 
critical exponents.  These lattices will also enable us to control the infinite-volume extrapolations in our future 
calculation for spectral quantities.

\section*{Acknowledgments}
We warmly thank Robert Shrock for helpful conversations.  This work is
supported by Taiwanese NSC via grants 100-2745-M-002-002-ASP
(Academic Summit Grant) and 99-2112-M-009-004-MY3,  the DFG through
the DFG-project Mu932/4-4, and the JSPS with Grant-in-Aid for Scientific Research (S)
number 22224003.  Simulations have been performed at the SGI system HLRN-II 
at the HLRN supercomputing service Berlin-Hannover,
the PAX cluster at DESY-Zeuthen, and HPC facilities at National
Chiao-Tung University and National Taiwan University.

\end{document}